\newcommand{\CLASSINPUToutersidemargin}{0.625in}
\documentclass[conference, romanappendices]{IEEEtran}
\usepackage{graphicx}
\usepackage{amsmath}
\usepackage{booktabs}
\usepackage{stmaryrd}
\usepackage{subfig}
\newtheorem{theorem}{Theorem}
\newtheorem{lemma}{Lemma}
\newtheorem{conjecture}{Conjecture}
\newtheorem{corollary}{Corollary}
\newtheorem{definition}{Definition}

\begin{document}

\title{Refining a Quantitative Information Flow Metric}

\author{\IEEEauthorblockN{Sari Haj Hussein\\}
\IEEEauthorblockA{Department of Computer Science, Aalborg University, Denmark\\
Email: angyjoe@gmail.com}}

\maketitle

\begin{abstract}
We introduce a new perspective into the field of quantitative information flow (QIF) analysis that invites the community to bound the leakage, reported by QIF quantifiers, by a range consistent with the size of a program's secret input instead of by a mathematically sound (but counter-intuitive) upper bound of that leakage. To substantiate our position, we present a refinement of a recent QIF metric that appears in the literature. Our refinement is based on slight changes we bring into the design of that metric. These changes do not affect the theoretical premises onto which the original metric is laid. However, they enable the natural association between flow results and the exhaustive search effort needed to uncover a program's secret information (or the residual secret part of that information) to be clearly established. The refinement we discuss in this paper validates our perspective and demonstrates its importance in the future design of QIF quantifiers.
\end{abstract}


\section{Introduction}\label{Section 1}
The goal of information flow analysis is to enforce limits on the use of information that apply to all computations that involve that information. For instance, a confidentiality property requires that a program with secret inputs should not leak those inputs into its public outputs. Qualitative information flow properties, such as non-interference are expensive, impossible, or rarely satisfied by real programs: generally some flow exists, and many systems remain secure provided that the amount of flow is sufficiently small, moreover, designers wish to distinguish acceptable from unacceptable flows.

Systems often reveal a summary of secret information they store. The summary contains fewer bits and provides a limit on the attacker's inference. For instance, a patient's report is released with the disease name covered by a black rectangle. However, it is not easy to precisely determine how much information exists in the summary. For instance, if the font size is uniform on the patient's report, the width of the black rectangle might determine the length of the disease name. Quantitative information flow (QIF) analysis is an approach that establishes bounds on information that is leaked by a program. In QIF, confidentiality properties are also expressed, but as limits on the number of bits that might be revealed from a program's execution. A violation is declared if the number of leaked bits exceeds the policy. Because information theory forms the foundation of QIF analysis, it should be possible to associate the quantities reported by QIF quantifiers with the effort needed to uncover secret information via exhaustive search. However, establishing this association is infeasible with QIF quantifiers that do not report a flow \emph{consistent} with the size of a program's secret input, but instead a mathematically sound \emph{upper bound} of that flow \cite{Clarkson2005}. For instance, consider the QIF metric and the password checker in Section $1$ of \cite{Clarkson2005}, and assume that the password space has a cardinality of $3$. This means that the size of the password is $\log3=1.5849$ bits. (Here and hereafter, all logarithms are to the base $2$). Nonetheless, the metric in \cite{Clarkson2005} might report a flow that exceeds $1.5849$ bits, which makes it impossible to determine the space of the exhaustive search that should be carried out in order to reveal the residual secret part of the password. However, if the flow reported is always less than $1.5849$ bits, the exhaustive search space becomes evident.

We believe that the counter-intuitive flow quantities reported by some QIF quantifiers, that appear in the literature, are due to a flaw in the design of those quantifiers, and that simple tweaks can bound those quantities by a range consistent with the size of a program's secret input. This paper takes the first step in this direction and refines the QIF metric suggested in \cite{Clarkson2005}. The metric in \cite{Clarkson2005} is based on a new perspective for QIF analysis. The fundamental idea is to model an attacker's belief about a program's secret input as a probability distribution over high states. This belief is then revised, using Bayesian updating techniques, as the attacker interacts with a program's execution. It is believed that the work reported in \cite{Clarkson2005} is the first to address an attacker's belief in quantifying information flow. This work was later expanded and appeared in \cite{Clarkson2009}. A number of relevant results \cite{Smith2009,Hamadou2010} were reported in the sequel; however, the work in \cite{Clarkson2005,Clarkson2009} is sufficient as a foundation of our work.

\subsection{Plan of the Paper}
The remainder of this paper is organized as follows. Section \ref{Section 2} elaborates on accuracy-based information flow analysis which is the major contribution in \cite{Clarkson2005}. In this section, we give concise elucidation of the elements of this analysis and how it differs from the classical uncertainty-based information flow analysis. In addition, we uncover some inexplicable results reported by the QIF metric in \cite{Clarkson2005}, and argue that the reasoning of this metric's designers is \emph{incomplete}. We further state the general range of flow reported by the metric in \cite{Clarkson2005} that applies to both deterministic and probabilistic programs as well as to all types of attacker's beliefs. This range is neither given in \cite{Clarkson2005} nor in \cite{Clarkson2009}. Over the course of acquiring the range, we reveal the ineffectiveness of the admissibility restriction suggested in \cite{Clarkson2005}. At the end of Section \ref{Section 2}, we conjecture a simple fix that can bound the results reported by the metric in \cite{Clarkson2005}. Underpinning our arguments in Section \ref{Section 2} is a formal definition of a \emph{size-consistent QIF quantifier}. Our definition is based on uncertainty-based information flow analysis, and it inaugurates the new perspective we are introducing into the field of QIF. To the best of our knowledge, this is the \emph{first} definition to capture the correlation between the size of a program's secret input and the quantification of flow from that input in the general case. Section \ref{Section 3} concentrates on Kullback-Leibler divergence which is a centerpiece of the metric in \cite{Clarkson2005}. We give some mathematical interpretations of this divergence, and then focus on its discrimination construct, suggesting the replacement of this construct with a better one, and subsequently the replacement of the divergence itself with another, bounded, divergence. This paves the way for the refinement of the metric in \cite{Clarkson2005} which is what we fulfill in two stages in Section \ref{Section 4}. We also give the range and the interpretation of the refined metric, and prove its properties and their meaningfulness compared to the original one, while minding the consistency of the probability distributions dealt with. Having justified the conjecture we made in Section \ref{Section 2}, and shown that a large number of possible refinements of the metric in \cite{Clarkson2005} exist, we discuss the association of the original and the refined metric with the exhaustive search effort in Section \ref{Section 5}, give some remarks in Section \ref{Section 6}, and conclude the paper in Section \ref{Section 7}. The proofs are given in Appendix \ref{Appendix I}.

\section{Uncertainty- vs. Accuracy-based Information Flow Analysis}\label{Section 2}
The problem with uncertainty-based information flow analysis is that it \emph{ignores} reality. As an example, consider a simple password checker $\mathcal{PWC}$ \cite{Clarkson2005} that sets an authentication flag $a$ after checking a stored password $p$ against a guessed password $g$ supplied by the user.
\begin{equation}\label{Equation 1}
    \mathcal{PWC}:\textnormal{ if }p=g\textnormal{ then }a:=1\textnormal{ else }a:=0
\end{equation}

For simplicity, suppose that the password space is $\mathcal{W}_{p}=\{A,B,C\}$, which gives a size of $\log|\mathcal{W}_{p}|=\log3=1.5849$ bits for the password $p$. Suppose further that the user is actually an attacker attempting to discover the password. \emph{Before} interacting with a $\mathcal{PWC}$ execution, this attacker believes that the password is \emph{overwhelmingly} likely to be $A$ but has a very small and equally likely chance to be either $B$ or $C$. More concretely and adopting the convention in \cite{Clarkson2005}, the attacker's \emph{prebelief} about $p$ is captured using a probability distribution $b_{H}:\mathcal{W}_{p}\rightarrow[0,1]$ as shown in Table \ref{Table 1}.

\begin{table}[h]
    \centering
    \subfloat[][Attacker's prebelief]{
        \begin{tabular}{lrrr}
        \toprule
        $p$ & $A$ & $B$ & $C$ \\
        \midrule
        $b_{H}$ & $0.98$ & $0.01$ & $0.01$ \\
        \bottomrule
        \end{tabular}
        \label{Table 1}
    }\qquad
    \subfloat[][Attacker's postbelief]{
        \begin{tabular}{lrrr}
        \toprule
        $p$ & $A$ & $B$ & $C$ \\
        \midrule
        $b_{H}^{'}$ & $0$ & $0.5$ & $0.5$ \\
        \bottomrule
        \end{tabular}
        \label{Table 2}
    }
    \caption{Attacker's beliefs in the password $p$}
\end{table}

The attacker's uncertainty about $p$ (\emph{not} necessarily about the correct $p$) is obtained via a simple application of Shannon uncertainty functional \cite{Halpern2003}:
\begin{equation*}
    \mathcal{U}=S(b_{H})=-0.98\log0.98-2\cdot0.01\log0.01=0.1614\textnormal{ bits}
\end{equation*}

Assuming that the correct password (the \emph{reality}) is $C$, if the attacker complies to her prebelief and feeds a $\mathcal{PWC}$ execution with $g=A$, \emph{she} will observe $a$ equal to $0$. The attacker then infers that $A$ is not the real password, and that there is an equal chance of $50\%$ that the password is either $B$ or $C$. As a result, the attacker's postbelief distributes as shown in Table \ref{Table 2}, and the attacker's uncertainty about $p$ becomes:
\begin{equation*}
    \mathcal{U}=S(b_{H}^{'})=-0.5\log0.5-0.5\log0.5=1\textnormal{ bit}
\end{equation*}

To complete an uncertainty-based information flow analysis, we have to compute the \emph{reduction} in uncertainty by subtracting the post- from the pre-uncertainty using the formula:
\begin{equation*}
    \mathcal{R}=\mathcal{U}-\mathcal{U}'
\end{equation*}

This gives us $\mathcal{R}=0.1614-1=-0.8386$ bits. In the sense of uncertainty-based analysis, the negative $\mathcal{R}$ means \emph{absence} of information flow. There is \emph{nothing} wrong with this interpretation provided that we do not connect information flow with how \emph{far} an attacker's belief is from reality. However, if we connect the flow with the distance between an attacker's belief and reality, then the interpretation that $\mathcal{R}$ supports does not make sense. The measure $\mathcal{R}$ ignores reality by measuring $b_{H}$ and $b_{H}^{'}$ against each other only, instead of against the high state (which is $C$ as the correct password in our example). It is good to notice however that the range of flow reported by $\mathcal{R}$ is as given by the formula:
\begin{equation*}
    \varrho_{\mathcal{R}}=[-\log|\mathcal{W}_{p}|,\log|\mathcal{W}_{p}|]=[-1.5849,1.5849]
\end{equation*}

This is a direct consequence of Shannon uncertainty functional falling in the range $[0,\log|\mathcal{W}_{p}|]$ \cite{Cover2006}. The range $\varrho_{\mathcal{R}}$ reported by $\mathcal{R}$ is \emph{plausible} if we remember that the size of the password $p$ is $1.5849$ bits. We would like to take time defining the size-consistent QIF quantifier.

\begin{definition}[Size-consistent QIF Quantifier]\label{Definition 1}
We say that a QIF quantifier is size-consistent if its reported results are bounded (from above and from below) by the size of a program's secret input. Formally, let $\mathcal{QUAN}$ be a QIF quantifier, and assume that the size of a program's secret input is $\eta$ bits. We say that $\mathcal{QUAN}$ is size-consistent if:
\begin{equation*}
    \mathcal{QUAN}_{max}\leq\eta\textnormal{ and }\mathcal{QUAN}_{min}\geq-\eta
\end{equation*}
\end{definition}

However, if we merely look at the attacker's prebelief and postbelief in $C$, as the correct password, we realize that the attacker's belief has \emph{approached} reality from interacting with $\mathcal{PWC}$. Approaching reality cannot happen unless the attacker \emph{learns} something from an amount of information $\mathcal{PWC}$ has conveyed. This conveyance corresponds to \emph{positive} information flow that informs the attacker, and flatly contradicts the uncertainty-based interpretation.

The earliest investigation of this \emph{specific} inadequacy of uncertainty-based information flow analysis appeared in \cite{Clarkson2005} and was later expanded in \cite{Clarkson2009}. The authors of \cite{Clarkson2009} propose to respect reality through what they call "accuracy-based information flow analysis". This sort of analysis has two elements:

\begin{enumerate}
  \item[E1.] Quantifying information flow from a program's execution to an attacker.
  \item[E2.] Respecting the distance between an attacker's belief and reality.
\end{enumerate}

The uncertainty-based analysis does \emph{not} have the second element as the example above demonstrated. The accuracy-based analysis quantifies flow as the \emph{improvement} in the accuracy of an attacker's belief. This is equivalent to saying the reduction in the distance between an attacker's belief and reality. The metric advanced in \cite{Clarkson2009} is based on this notion of improvement, and is given by the formula:
\begin{equation}\label{Equation 2}
    \mathcal{Q}(\mathcal{E},b_{H}^{'})=D(b_{H}\rightarrow\dot{\sigma}_{H})-D(b'_{H}\rightarrow\dot{\sigma}_{H})
\end{equation}
where $\mathcal{E}=\langle S,b_{H},\sigma_{H},\sigma_{L}\rangle$ is an experiment tuple as defined in \cite{Clarkson2009}, $\langle\mathcal{E},b_{H}^{'}\rangle$ is the outcome of that experiment, $b_{H}$ is the attacker's prebelief, $b_{H}^{'}$ is the attacker's postbelief, $\dot{\sigma}_{H}$ is a probability distribution that maps the high state $\sigma_{H}$ to $1$ (this is the \emph{certainty} about the high state; about \emph{reality}), and $D$ is Kullback-Leibler divergence (also known as relative entropy or information gain \cite{Cover2006}) given by the formula:
\begin{equation}\label{Equation 3}
    D(b\rightarrow b')=\underset{\sigma\in\mathcal{W}_{p}}{\sum}b'(\sigma)\cdot\log\frac{b'(\sigma)}{b(\sigma)}
\end{equation}

Notice in formula (\ref{Equation 2}) how $\mathcal{Q}$ respects reality by measuring $b_{H}$ and $b_{H}^{'}$ against the correct high state $\dot{\sigma}_{H}$, instead of against each other only. Formula (\ref{Equation 2}) is simplified in \cite{Clarkson2009} to (this simplification is \emph{reality-aware}):
\begin{equation}\label{Equation 4}
    \begin{array}{l}
    \mathcal{Q}(\mathcal{E},b_{H}^{'})=D(b_{H}\rightarrow\dot{\sigma}_{H})-D(b_{H}^{'}\rightarrow\dot{\sigma}_{H})\\
    \hphantom{\mathcal{Q}(\mathcal{E},b_{H}^{'})}=\underset{\sigma\in\mathcal{W}_{p}}{\sum}\dot{\sigma}_{H}(\sigma)\cdot\log\frac{\dot{\sigma}_{H}(\sigma)}{b_{H}(\sigma)}\\
    \hphantom{\mathcal{Q}(\mathcal{E},b_{H}^{'})}-\underset{\sigma\in\mathcal{W}_{p}}{\sum}\dot{\sigma}_{H}(\sigma)\cdot\log\frac{\dot{\sigma}_{H}(\sigma)}{b_{H}^{'}(\sigma)}\\
    \hphantom{\mathcal{Q}(\mathcal{E},b_{H}^{'})}=-\log b_{H}(\sigma_{H})+\log b_{H}^{'}(\sigma_{H})
    \end{array}
\end{equation}

To complete an accuracy-based information flow analysis parallel to the uncertainty-based analysis we have completed earlier in this section, we apply formula (\ref{Equation 4}) to the same example given above to obtain:
\begin{equation}\label{Equation 5}
    \mathcal{Q}(\mathcal{E},b_{H}^{'})=-\log0.01+\log0.5=5.6438\textnormal{ bits}
\end{equation}

The flow value of $5.6438$ bits reported by $\mathcal{Q}$ violates the plausible range $\varrho_{\mathcal{R}}=[-1.5849,1.5849]$ and equally exceeds the size needed to store the password $p$. How can a flow from $p$ exceed the size needed to store $p$? A sound but puzzling result in the \emph{field} of QIF analysis that the authors of \cite{Clarkson2009} attribute to that the attacker's prebelief is not uniform; it is more erroneous than a \emph{uniform} belief ascribing $1/3$ probability to each password $A$, $B$, and $C$, and therefore a \emph{larger} amount of information is required to correct it! But what can the source of this larger amount of information be? Is it a \emph{covert} agent external to the system and the attacker when all the agents are assumed condensed to just the attacker and the system \cite{Clarkson2009}? Besides is it always true that a uniform attacker's prebelief would, in a series of experiments, cause her to learn a total of $\log3$ bits \cite{Clarkson2009}? This claim is valid for a deterministic password checker, but \emph{incomplete} for a probabilistic one. Let us verify this fact.

It is proved in \cite{Clarkson2009} that for deterministic programs (including the deterministic $\mathcal{PWC}$ given in formula (\ref{Equation 1})), we have:
\begin{equation}\label{Equation 6}
    b_{H}(\sigma_{H})\leq b_{H}^{'}(\sigma_{H})
\end{equation}
Since $b_{H}^{'}$ is a probability distribution, we can write:
\begin{equation*}
    b_{H}(\sigma_{H})\leq b_{H}^{'}(\sigma_{H})\leq1
\end{equation*}
which means:
\begin{equation*}
    \begin{array}{c}
    \log b_{H}(\sigma_{H})\leq\log b_{H}^{'}(\sigma_{H})\leq0\\
    0\leq\mathcal{Q}\leq-\log b_{H}(\sigma_{H})
    \end{array}
\end{equation*}
The attacker's prebelief is assumed uniform on $\mathcal{W}_{p}$, therefore:
\begin{equation*}
    0\leq\mathcal{Q}\leq\log3
\end{equation*}

Thus, it is beyond a shadow of a doubt that a uniform attacker's prebelief would cause her to learn a total of $\log3$ bits from interacting with a deterministic $\mathcal{PWC}$. But does the attacker's learning outcome differ when interacting with a probabilistic $\mathcal{PWC}$? An illustrative probabilistic $\mathcal{PWC}$ is:
\begin{equation*}
    \begin{array}{l}
    \mathcal{PPWC}:\textnormal{ if }p=g\textnormal{ then }a:=1\,_{0.99}\talloblong\, a:=0\\
    \hphantom{\mathcal{PPWC}:\textnormal{ if }p=g}\textnormal{ else }a:=0\,_{0.99}\talloblong\, a:=1
    \end{array}
\end{equation*}

The inequality in formula (\ref{Equation 6}) no longer holds, and we are free to write:
\begin{equation*}
    \begin{array}{c}
    0\leq b_{H}^{'}(\sigma_{H})\leq1\\
    -\infty\leq-\log b_{H}(\sigma_{H})+\log b_{H}^{'}(\sigma_{H})\leq-\log b_{H}(\sigma_{H})\\
    -\infty\leq Q\leq0\textnormal{ or }0\leq Q\leq\log3
    \end{array}
\end{equation*}

The sub-range $-\infty\leq Q\leq0$ shows that a uniform attacker's prebelief might cause her to learn an infinite number of \emph{misinforming} bits from interacting with $\mathcal{PPWC}$. This demonstrates the \emph{incompleteness} of the claim "a uniform attacker's prebelief would, in a series of experiments, cause her to learn a total of $\log3$ bits" made in \cite{Clarkson2009}.

The previous discussion motivates the investigation of the general range of the $\mathcal{Q}$ metric that holds with both deterministic and probabilistic programs as well as with all types of attacker's beliefs. This range is attained in Lemma \ref{Lemma 1}.

\begin{lemma}\label{Lemma 1}
Considering both deterministic and probabilistic programs, and all types of an attacker's beliefs, the general range of flow reported by $\mathcal{Q}$ is:
\begin{equation*}
    \varrho_{\mathcal{Q}}=(-\infty,-\log b_{H}(\sigma_{H})]
\end{equation*}
\end{lemma}

Clearly $\mathcal{Q}$ is not size-consistent. Let us now muse on the computation in formula (\ref{Equation 5}) and try to figure out a mean to proceed with this correspondence. The flow of $5.6438$ bits has brought the attacker from $-\log0.01=6.6438$ bits away from reality to $-\log0.5=1$ bits away from it. In addition and as proved in Theorem $3$ in \cite{Clarkson2009}, each bit of flow has made the attacker twice as likely to guess \emph{correctly} \cite{Massey1994}, or equivalently twice as certain about the \emph{correct} high state (in total, we have $2^{5.6438}\approx50$ times increase in the likelihood of a correct guess). In the uncertainty-based definition, the attacker's certainty is ascribed to a high state that \emph{might} be incorrect...Conjecture \ref{Conjecture 1} engrossedly stops the correspondence.

\begin{conjecture}\label{Conjecture 1}
Considering Theorem $3$ in \cite{Clarkson2009}, if a bit of flow makes the attacker \emph{more} than twice as likely to guess correctly, then $\mathcal{Q}$ should become size-consistent.
\end{conjecture}

Seeking a justification for this conjecture will be the purpose of the later sections. Although the authors of \cite{Clarkson2005,Clarkson2009} are acclaimed for their contribution to the field of QIF through their accuracy-based analysis, their metric allows the respect for reality (element E2) to attenuate the quality of flow quantification (element E1). This attenuation is the result of \emph{severe} discrimination in Kullback-Leibler divergence as we shall see in the next section.

\section{Concentrating on Kullback-Leibler Divergence}\label{Section 3}
\subsection{Possible Interpretations of the Divergence}
The divergence $D$ between $b$ and $b'$, given in formula (\ref{Equation 3}), can be interpreted in terms of code \emph{inefficiency} as follows; $D$ is the average number of bits that are wasted by encoding events from a distribution $b'$ with a code based on a not-quite-right distribution $b$ \cite{Manning1999}. Another way of writing $D$ in terms of the \emph{expected} value function \cite{Klir2005} is as follows:
\begin{equation*}
    D(b\rightarrow b')=E_{b'}(\log\frac{b'(\sigma)}{b(\sigma)}),\; E_{b'}(f)=\underset{\sigma\in\mathcal{W}_{p}}{\sum}b'(\sigma)\cdot f(\sigma)
\end{equation*}

The function $E_{b'}$ takes the weighted average of the values $f(\sigma)$ in which the weights are probabilities $b'$. In the original paper by Kullback and Leibler \cite{Kullback1951}, the values:
\begin{equation}\label{Equation 7}
    \mathcal{I}_{Dis}(\sigma)=\log\frac{b'(\sigma)}{b(\sigma)}
\end{equation}
are seen as the information in $\sigma$ for the \emph{discrimination} between $b$ and $b'$. This is plausible if we rewrite the previous values as:
\begin{equation*}
    -\log b(\sigma)-(-\log b'(\sigma))
\end{equation*}
and recall that the information contained in an observation of an event $E$ with probability $p(E)$ is $-\log p(E)$ \cite{Cover2006}.

This notion of discrimination leads to another interpretation of $D$; it is the weighted average of the information in $\sigma$ for the discrimination between $b$ and $b'$ where the weights are probabilities $b'$. We write:
\begin{equation}\label{Equation 8}
    D(b\rightarrow b')=E_{b'}(\mathcal{I}_{Dis}(\sigma))
\end{equation}

\subsection{A Better Discrimination Construct}
We propose to replace the discrimination construct in formula (\ref{Equation 8}) with the following:
\begin{equation}\label{Equation 9}
    \mathcal{I}_{Dis}^{'}(\sigma)=\log\frac{b'(\sigma)}{\frac{b'(\sigma)+b(\sigma)}{2}}
\end{equation}
for $\mathcal{I}_{Dis}^{'}(\sigma)$ to be the information in $\sigma$ for the discrimination between the mean $(b'+b)/2$ and $b'$. But what is the effect of this replacement? The following lemma shows that we have actually cut down the discrimination \emph{at least} by half.

\begin{lemma}\label{Lemma 2}
The proposed discrimination construct cuts down the discrimination in Kullback-Leibler divergence at least by half, that is: $\mathcal{I}_{Dis}^{'}(\sigma)\leq\frac{1}{2}\mathcal{I}_{Dis}(\sigma)$.
\end{lemma}

\begin{figure*}
    \centering
    {\small
    \hfill{}
    \subfloat[][Between $\mathcal{I}_{Dis}(\sigma)$ and $\mathcal{I}_{Dis}^{'}(\sigma)$]{
        \includegraphics[scale=0.25]{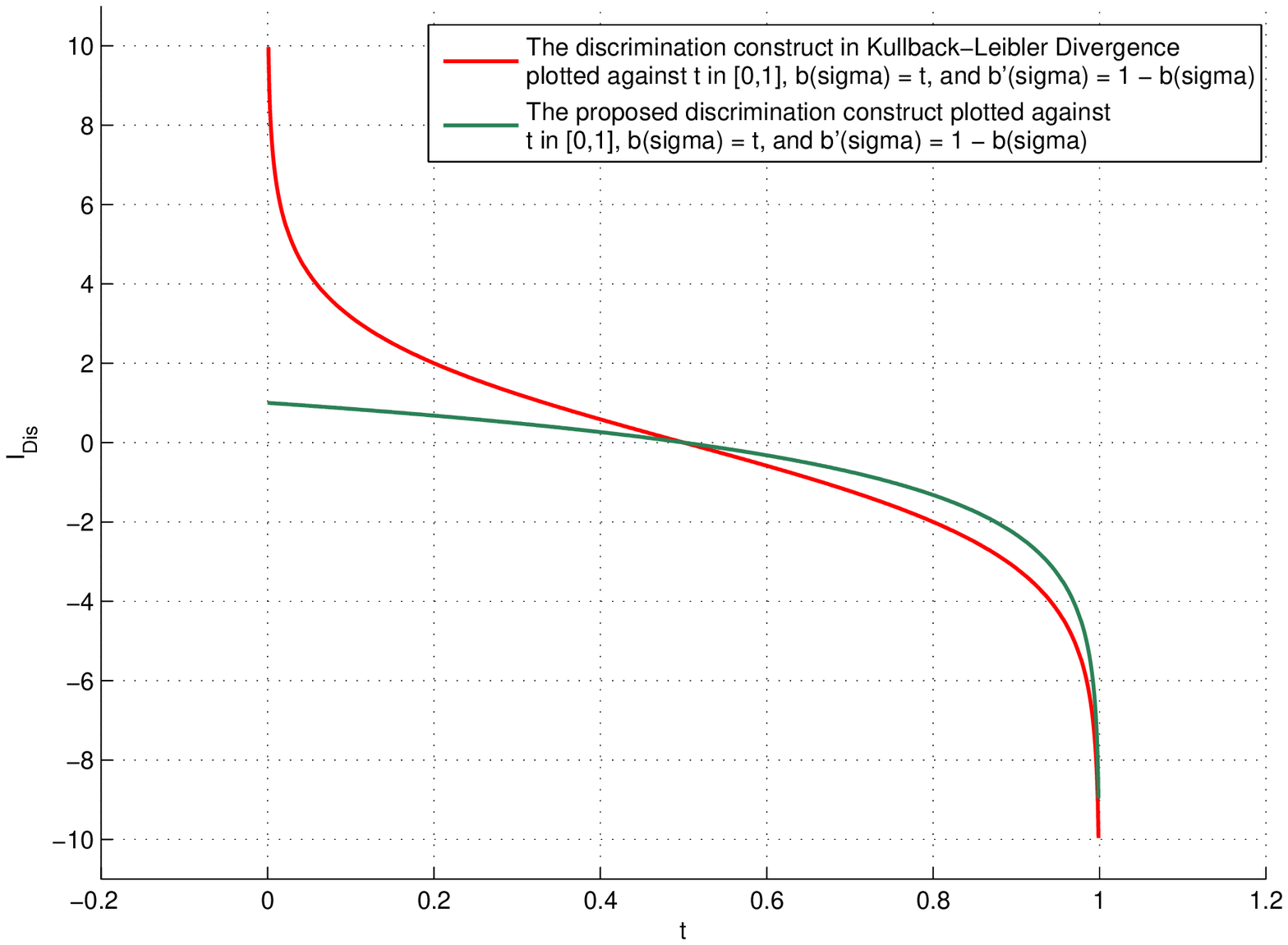}
        \label{Figure 1}
    }\qquad
    \subfloat[][Between $D$ and $D'$]{
        \includegraphics[scale=0.25]{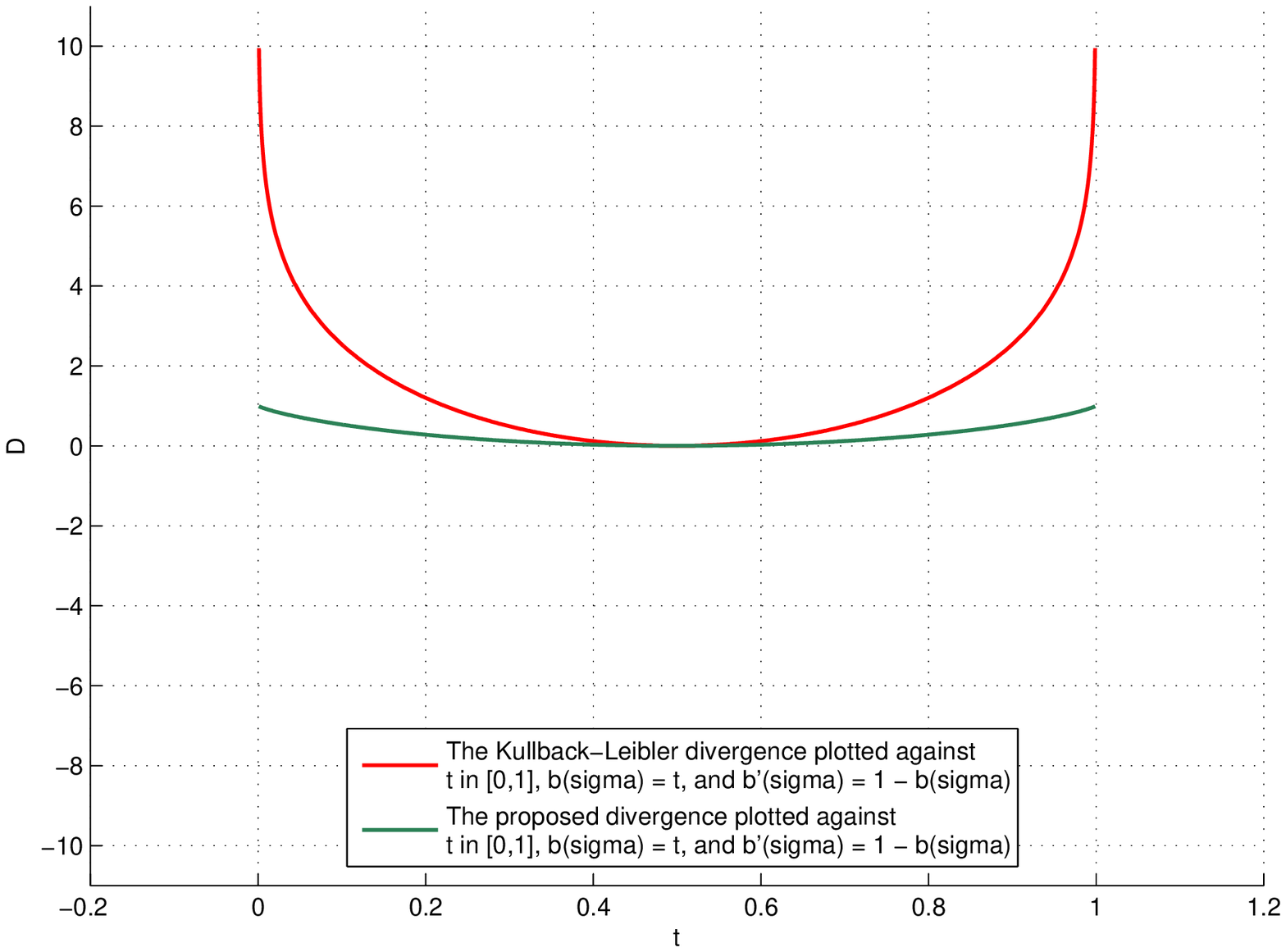}
        \label{Figure 2}
    }\qquad
    \subfloat[][Between $\mathcal{Q}$ and $\mathcal{Q}''$]{
        \includegraphics[scale=0.25]{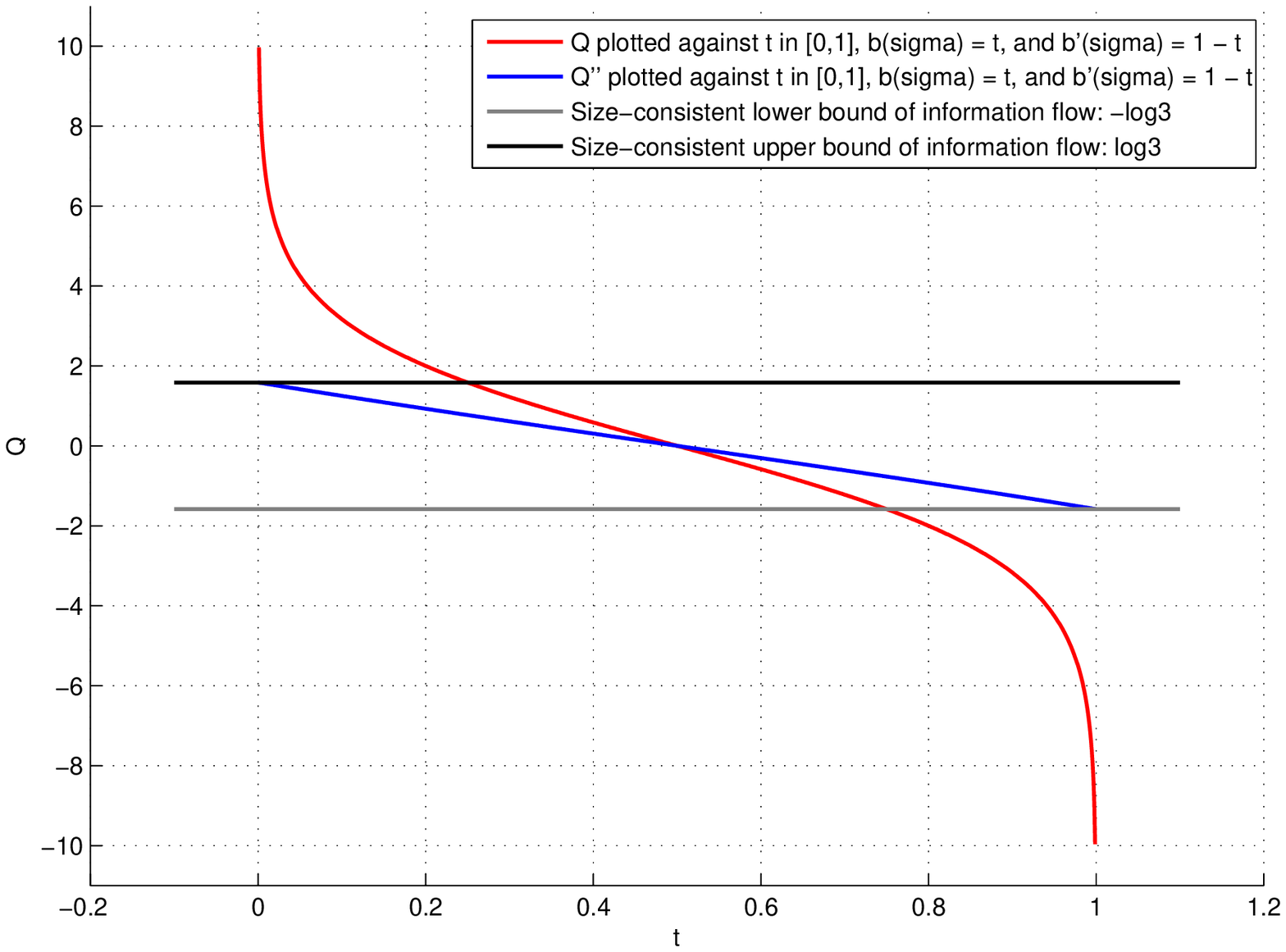}
        \label{Figure 3}
    }
    \caption{Graphical comparisons made in the paper}}
\end{figure*}

A graphical comparison between $\mathcal{I}_{Dis}(\sigma)$ and $\mathcal{I}_{Dis}^{'}(\sigma)$ is shown in Figure \ref{Figure 1}. It is important to notice at this stage that halving the infinite value of $\mathcal{I}_{Dis}(\sigma)$ does \emph{not} make it finite.

\subsection{A Better Divergence}
Substituting (\ref{Equation 9}) for (\ref{Equation 7}) in (\ref{Equation 8}), we get the divergence:
\begin{equation}\label{Equation 10}
    D'(b\rightarrow b')=\underset{\sigma\in\mathcal{W}_{p}}{\sum}b'(\sigma)\cdot\log\frac{b'(\sigma)}{\frac{b'(\sigma)+b(\sigma)}{2}}
\end{equation}

The resulted divergence meets with the asymmetric form $K$ of Jensen-Shannon divergence proposed in \cite{Lin1991}. In fact, formula (\ref{Equation 9}) and Lemma \ref{Lemma 2} both appear in \cite{Lin1991} wrapped in the expected value function. $D'$ is nonnegative and equals zero if and only if $b=b'$ \cite{Lin1991}. This is essential for any measure of difference and \emph{justifies} using $D'$ instead of $D$ to measure the distance between two beliefs.
A possible interpretation of $D'$ is as follows; how much information is \emph{lost} if we describe the two random variables that correspond to $b$ and $b'$ with their average distribution $(b'+b)/2$? This interpretation gives $D'$ the nickname "information radius" \cite{Manning1999}.

A graphical comparison between $D$ and $D'$ is shown in Figure \ref{Figure 2}. Notice that $D$ approaches infinity when $t$ approaches $0$ or $1$. In contrast, $D'$ is always well defined in the entire range $t\in[0,1]$. This is because $(b'+b)/2\neq0$ if either $b'=0$ or $b=0$. But what is the effect of using $D'$ instead of $D$ in $\mathcal{Q}$? This will be our focus in the next section.

\section{Refining the Metric}\label{Section 4}

\subsection{Refining to Normalization}
If we substitute (\ref{Equation 10}) for (\ref{Equation 3}) in (\ref{Equation 2}), we get the metric:
\begin{equation*}
    \begin{array}{l}
    \mathcal{Q}'(\mathcal{E},b_{H}^{'})=D'(b_{H}\rightarrow\dot{\sigma}_{H})-D'(b_{H}^{'}\rightarrow\dot{\sigma}_{H})\\
    \hphantom{\mathcal{Q}'(\mathcal{E},b_{H}^{'})}=\underset{\sigma\in\mathcal{W}_{p}}{\sum}\dot{\sigma}_{H}(\sigma)\cdot\log\frac{\dot{\sigma}_{H}(\sigma)}{\frac{\dot{\sigma}_{H}(\sigma)+b_{H}(\sigma)}{2}}\\
    \hphantom{\mathcal{Q}'(\mathcal{E},b_{H}^{'})}-\underset{\sigma\in\mathcal{W}_{p}}{\sum}\dot{\sigma}_{H}(\sigma)\cdot\log\frac{\dot{\sigma}_{H}(\sigma)}{\frac{\dot{\sigma}_{H}(\sigma)+b_{H}^{'}(\sigma)}{2}}\\
    \hphantom{\mathcal{Q}'(\mathcal{E},b_{H}^{'})}=-\log(1+b_{H}(\sigma_{H}))+\log(1+b_{H}^{'}(\sigma_{H}))
    \end{array}
\end{equation*}

Notice that the above substitution does \emph{not} destroy the bedrock of accuracy-based analysis which, as mention in Section \ref{Section 2}, quantifies flow as the improvement in the accuracy of an attacker's belief. This \emph{guarantees} that $\mathcal{Q}'$ is a real metric of information flow. Before proceeding any further, we need to investigate the general range of $\mathcal{Q}'$, which is what we do in Lemma \ref{Lemma 3}.

\begin{lemma}\label{Lemma 3}
Considering both deterministic and probabilistic programs, and all types of an attacker's beliefs, and \emph{avoiding} the imposition of any admissibility restriction on those beliefs, the general range of flow reported by $\mathcal{Q}'$ is:
\begin{equation*}
    \varrho_{\mathcal{Q}'}=[-1,1]
\end{equation*}
\end{lemma}

Fortunately, the sub-range $[-1,0]$ corresponds to the attacker's \emph{misinformation} while the sub-range $[0,1]$ corresponds to the attacker's \emph{information} about the correct high state.

The new range $\varrho_{\mathcal{Q}'}=[-1,1]$, we have reached, does not make $\mathcal{Q}'$ size-consistent. Nonetheless, $\varrho_{\mathcal{Q}'}$ is a plausible normalization (flow \emph{percentage}) that is invariant with respect to the choice of the measurement unit.

\subsection{Refining to Actuality}
To ensure bits as the measurement unit, and avoid the need to transform the flow results back and forth between the ranges $\varrho_{\mathcal{Q}'}=[-1,1]$ and $\varrho_{\mathcal{R}}=[-1.5849,1.5849]$, we let $\eta$ be the size of a program's secret input in bits, and define the refined metric as:
\begin{equation}\label{Equation 11}
    \begin{array}{l}
    \mathcal{Q}''(\mathcal{E},b_{H}^{'})=\eta\cdot\mathcal{Q}'(\mathcal{E},b_{H}^{'})\\
    \hphantom{\mathcal{Q}''(\mathcal{E},b_{H}^{'})}=\eta\cdot[-\log(1+b_{H}(\sigma_{H}))+\log(1+b_{H}^{'}(\sigma_{H}))]
    \end{array}
\end{equation}

A graphical comparison between $\mathcal{Q}$ and $\mathcal{Q}''$ in the case of $\mathcal{PWC}$, along with the size-consistent uncertainty-based upper and lower bounds of flow, is shown in Figure \ref{Figure 3}. It is important to notice in this figure that the parts of the $\mathcal{Q}$ and $\mathcal{Q}''$ graphs that fall \emph{above} the zero mark on the $Y$ axis represent the attacker's information about the correct high state. In contrast, the attacker's misinformation is represented by the parts that fall \emph{below} the zero mark on the $Y$ axis. Another important observation to make in this figure is that, akin to $\mathcal{Q}$, $\mathcal{Q}''$ is \emph{sensitive} to changes in the attacker's belief. It is thus noted that $\mathcal{Q}''$ is a good quantifier of flow (element E1) that adheres well to reality (element E2).

\subsection{Range of the Refined Metric}
The most celebrated property of the refined metric is probably its range which is sought in Theorem \ref{Theorem 1}.
\begin{theorem}\label{Theorem 1}
Considering both deterministic and probabilistic programs, and all types of an attacker's beliefs, and \emph{avoiding} the imposition of any admissibility restriction on those beliefs, the general range of flow reported by $\mathcal{Q}''$ is:
\begin{equation*}
    \varrho_{\mathcal{Q}''}=[-\eta\cdot\log(1+b_{H}(\sigma_{H})),\eta\cdot[1-\log(1+b_{H}(\sigma_{H}))]]
\end{equation*}
where $\eta$ is the size of a program's secret input in bits.
\end{theorem}

\begin{corollary}\label{Corollary 1}
Notice that $\log(1+b_{H}(\sigma_{H}))\leq1$. This means that $\mathcal{Q}_{max}^{''}\leq\eta$ and $\mathcal{Q}_{min}^{''}\geq-\eta$, and makes $\mathcal{Q}''$ size-consistent.
\end{corollary}

\subsection{Interpreting the Refined Metric}
If we apply formula (\ref{Equation 11}) to the same example given in Section \ref{Section 2}, we get:
\begin{equation*}
    \mathcal{Q}''(\mathcal{E},b_{H}^{'})=0.9044\textnormal{ bits}
\end{equation*}

This time, the flow of $0.9044$ bits has brought the attacker from
\begin{equation*}
    1.5849\cdot[1-\log(1+0.01)]=1.5621\textnormal{ bits}
\end{equation*}
away from reality to
\begin{equation*}
    1.5849\cdot[1-\log(1+0.5)]=0.6577\textnormal{ bits}
\end{equation*}
away from it. But how much did this flow make the attacker likely to guess correctly? Theorem \ref{Theorem 2} answers this question, substantiating the validity of Conjecture \ref{Conjecture 1} we made in Section \ref{Section 2}, and showing that a bit of flow reported by $\mathcal{Q}''$ makes the attacker \emph{more} than twice as likely to guess correctly.

\begin{theorem}\label{Theorem 2}
A flow of $k$ bits reported by $\mathcal{Q}''$ makes the attacker \emph{more} than $2^{k}$ as likely to guess correctly. Strictly speaking:
\begin{equation}\label{Equation 12}
    \mathcal{Q}''(\mathcal{E},b_{H}^{'})=k\Leftrightarrow b_{H}^{'}(\sigma_{H})=2^{k/\eta}\cdot b_{H}(\sigma_{H})+2^{k/\eta}-1
\end{equation}
where $\eta$ is the size of a program's secret input in bits.
\end{theorem}

\subsection{Consistency of the Probability Distributions}
The bounds of $\mathcal{Q}''$, given in Theorem \ref{Theorem 1}, ensure \emph{proper} bounds of $b_{H}^{'}$. This can be easily shown by assuming a flow of $k$ bits and proceeding as follows:
\begin{equation*}
    \begin{array}{c}
    -\eta\cdot\log(1+b_{H}(\sigma_{H}))\leq k\leq\eta\cdot[1-\log(1+b_{H}(\sigma_{H}))]\\
    2^{\log(\frac{1}{1+b_{H}(\sigma_{H})})}\cdot(1+b_{H}(\sigma_{H}))-1\leq b_{H}^{'}(\sigma_{H})\\
    \leq2^{\log(\frac{2}{1+b_{H}(\sigma_{H})})}\cdot(1+b_{H}(\sigma_{H}))-1\\
    0\leq b_{H}^{'}(\sigma_{H})\leq1
    \end{array}
\end{equation*}

However, this does \emph{not} ensure that an intermediate value of $\mathcal{Q}''$ leads to $b_{H}^{'}$ falling outside the range $[0,1]$. To ensure this, we need to show that $\mathcal{Q}''$ is a monotone function. This is done in Lemma \ref{Lemma 4}.

\begin{lemma}\label{Lemma 4}
$\mathcal{Q}''$ is a monotonically increasing function, that is:
\begin{equation*}
    \forall b_{1},b_{2}:b_{1}\leq b_{2}\Rightarrow\mathcal{Q}''(\mathcal{E},b_{1})\leq\mathcal{Q}''(\mathcal{E},b_{2})
\end{equation*}
\end{lemma}

Thus, the probability distributions dealt with are invariably consistent.

\subsection{Meaningfulness of the Bounds}
We still have to accentuate the meaningfulness of the bounds of $\mathcal{Q}''$ in relation to the attacker's likelihood of a correct guess, or equivalently, to the attacker's certainty about the correct high state. This is done in Theorems \ref{Theorem 3} and \ref{Theorem 4}.

\begin{theorem}\label{Theorem 3}
An \emph{informing} flow equal to the upper bound of $\mathcal{Q}''$ is sufficient to make a fully \emph{uncertain} attacker fully \emph{certain} about the correct high state.
\end{theorem}

\begin{corollary}\label{Corollary 2}
Notice that, in the case of a fully uncertain attacker, we have:
\begin{equation*}
    \mathcal{Q}_{min}^{''}(\mathcal{E},b_{H}^{'})=-\eta\cdot\log(1+b_{H}(\sigma_{H}))=-\eta\cdot\log1=0
\end{equation*}
This yields the absolute range $\varrho_{\mathcal{Q}''}=[0,\eta]$ for $\mathcal{Q}''$, and reflects the rationality that a fully uncertain attacker can \emph{only} be informed.
\end{corollary}

\begin{theorem}\label{Theorem 4}
A \emph{misinforming} flow equal to the lower bound of $\mathcal{Q}''$ is sufficient to make a fully \emph{certain} attacker fully \emph{uncertain} about the correct high state.
\end{theorem}

A similar corollary to Corollary \ref{Corollary 2} can be stated to show that a fully certain attacker can \emph{only} be misinformed.

\subsection{Other Refinements}
The discrimination construct, given in formula (\ref{Equation 9}), which we used in our refinement is definitely \emph{not} the only apt construct. \emph{Any} construct that reduces the discrimination is a likely candidate for the replacement of the Kullback-Leibler construct (given in formula (\ref{Equation 7})). For instance, consider the following discrimination construct:
\begin{equation*}
    \mathcal{I}_{Dis}^{''}(\sigma)=\log\frac{1+b'(\sigma)}{1+b(\sigma)}
\end{equation*}

This construct clearly cuts down the discrimination. Moreover, it leads to the same refinement that the construct in (\ref{Equation 9}) had led to. This shows that there is a \emph{large} number of possible refinements of the $\mathcal{Q}$ metric. However, we favored the construct in (\ref{Equation 9}) since the properties of Jensen-Shannon divergence are well-examined in the literature \cite{Lin1991}.

\section{Exhaustive Search Effort}\label{Section 5}
Assuming a program with a secret input of size $\eta$ bits, and an informing flow of $k$ bits from the same program to an attacker. The dynamic upper bound of $\mathcal{Q}''$, given in Theorem \ref{Theorem 1}, tells us that $k\leq\eta$. Therefore, the space of the exhaustive search \cite{Menezes1997} that should be carried out in order to reveal the residual part $\eta-k$ bits of the secret input is $2^{\eta-k}$. On the other hand, the dynamic upper bound of $\mathcal{Q}$, given in Lemma \ref{Lemma 1}, tells us that $k>\eta$ is a possible scenario. In scenarios as such, the residual part of the secret input is impossible to determine, and consequently, the exhaustive search space cannot be established, albeit that the secret input might have been partially revealed to the attacker (refer to the example in Section \ref{Section 2}).

\section{Remarks}\label{Section 6}
In addition to the divergence $K$, given in formula (\ref{Equation 10}), Lin \cite{Lin1991} identified two other divergence measures. The first divergence is denoted as $J$, and is given by the formula:
\begin{equation*}
    J(b\rightarrow b')=\underset{\sigma\in\mathcal{W}_{p}}{\sum}(b'(\sigma)-b(\sigma))\cdot\log\frac{b'(\sigma)}{b(\sigma)}
\end{equation*}

This divergence is the symmetric form of Kullback-Leibler divergence, given in formula (\ref{Equation 3}), and they both share the same problems; they are unbounded from above and undefined if $b(\sigma)=0$ and $b'(\sigma)\neq0$ for any $\sigma\in\mathcal{W}_{p}$. It is therefore doubtful that the use of any of these two divergence measures would lead to size-consistent QIF quantifiers. The second divergence Lin identified is denoted as $L$, and is given by the formula:
\begin{equation*}
    L(b\rightarrow b')=2S(\frac{b+b'}{2})-S(b)-S(b')
\end{equation*}
where $S$ is Shannon uncertainty functional \cite{Halpern2003}. This divergence is the symmetric form of the divergence $K$ we used in our refinement. It has an obvious information-theoretic interpretation in terms of Shannon uncertainty functional which makes it suitable for use in accuracy-based information flow analysis when an attacker's belief about a program's secret input is modeled using advanced representations of uncertainty other than a simple probability distribution over high states. We leave the investigation of this use as future work.

\section{Conclusions}\label{Section 7}
We presented a refinement of the QIF metric in \cite{Clarkson2005,Clarkson2009} that bounds its reported results by a plausible range. Both the original and the refined metric are justified quantifiers of the flow that occurred during a program's execution. However, they differ in their interpretation of one bit of flow. Contrary to the original metric, the results reported by the refined metric are easily associated with the exhaustive search effort needed to uncover a program's secret information (or the residual secret part of that information). We believe that the counter-intuitive flow quantities reported by some QIF quantifiers, that appear in the literature, are due to a flaw in the design of those quantifiers. We further believe that this can be avoided by introducing minor changes into the design of those quantifiers.

\section*{Acknowledgment}
The author would like to thank Peter Y. A. Ryan and Marc Pouly for their helpful comments on an early draft of this paper.

\bibliographystyle{IEEEtran}
\bibliography{IEEEabrv,refining}

\appendices
\section{Proofs}\label{Appendix I}

\subsection{Proof of Lemma \ref{Lemma 1}}\label{Appendix A}
\noindent Kullback-Leibler divergence given in formula (\ref{Equation 3}) has the range:
\begin{equation*}
    0\leq D(b\rightarrow b')\leq+\infty
\end{equation*}
which means that:
\begin{equation*}
    \begin{array}{c}
    -\infty\leq D(b_{H}\rightarrow\dot{\sigma}_{H})-D(b'_{H}\rightarrow\dot{\sigma}_{H})\leq+\infty\\
    -\infty\leq\mathcal{Q}\leq+\infty
    \end{array}
\end{equation*}
It could be safer to bring the reader around by showing the \emph{extreme} cases. The extreme case from above $\mathcal{Q}=+\infty$ is reached when $b_{H}(\sigma_{H})=0$ and $b_{H}^{'}(\sigma_{H})=1$, whereas the converse yields the extreme case from below $\mathcal{Q}=-\infty$.
An admissibility restriction is suggested in \cite{Clarkson2005} on the attacker's prebelief. This restriction ensures that the prebelief never \emph{deviates} by more than a positive factor from a uniform distribution, and is given by the formula:
\begin{equation*}
    \textnormal{min}_{\sigma_{H}}(b_{H}(\sigma_{H}))\geq\epsilon\cdot\frac{1}{|\textnormal{State}_{H}|};\epsilon>0
\end{equation*}
The restriction above more or less excludes the attacker's initial belief that certain states are \emph{impossible}, or in other words, ascribing zero as a prebelief. However, it does \emph{not} impose anything on the attacker's postbelief, which enables us to write:
\begin{equation*}
    0<b_{H}(\sigma_{H}))\leq1\textnormal{ and }0\leq b_{H}^{'}(\sigma_{H})\leq1
\end{equation*}
and consequently:
\begin{equation*}
    \begin{array}{c}
    -\infty\leq D(b_{H}\rightarrow\dot{\sigma}_{H})-D(b'_{H}\rightarrow\dot{\sigma}_{H})<+\infty\\
    -\infty\leq\mathcal{Q}<+\infty
    \end{array}
\end{equation*}
Notice how the admissibility restriction is \emph{weak} in that it averts reporting infinite \emph{informing} flow
from the metric $\mathcal{Q}$, while leaving the rest of the counter-intuitive results unattended (perhaps this explains why the admissibility restriction is given in the original work \cite{Clarkson2005}, but not in the expanded one \cite{Clarkson2009}). We have yet to arrive at the general range of $\mathcal{Q}$. The last word on this matter relates to the fact that the attacker's postbelief about the correct high state can neither be better than full \emph{certainty} nor worse than full \emph{uncertainty}.
The former of these two arguments yields the \emph{dynamic} upper bound of $\mathcal{Q}$ which corresponds to the maximum \emph{informing} flow:
\begin{equation*}
    \mathcal{Q}_{max}(\mathcal{E},b_{H}^{'})=-\log b_{H}(\sigma_{H})+\log1=-\log b_{H}(\sigma_{H})
\end{equation*}
whereas the latter of the two arguments yields the \emph{absolute} lower bound of $\mathcal{Q}$ which corresponds to the maximum \emph{misinforming} flow:
\begin{equation*}
    \mathcal{Q}_{min}(\mathcal{E},b_{H}^{'})=-\log b_{H}(\sigma_{H})+\log0=-\infty
\end{equation*}
This gives us the general range of flow reported by $\mathcal{Q}$:
\begin{equation*}
    \varrho_{\mathcal{Q}}=(-\infty,-\log b_{H}(\sigma_{H})]
\end{equation*}

\subsection{Proof of Lemma \ref{Lemma 2}}\label{Appendix B}
\noindent The inequality of the arithmetic and geometric means gives us:
\begin{equation*}
    \frac{b'(\sigma)+b(\sigma)}{2}\geq\sqrt{b'(\sigma)\cdot b(\sigma)}
\end{equation*}
Based on this, we can write:
\begin{equation*}
    \mathcal{I}_{Dis}^{'}(\sigma)=\log\frac{b'(\sigma)}{\frac{b'(\sigma)+b(\sigma)}{2}}\leq\log\frac{b'(\sigma)}{\sqrt{b'(\sigma)\cdot b(\sigma)}}=\frac{1}{2}\mathcal{I}_{Dis}(\sigma)
\end{equation*}

\subsection{Proof of Lemma \ref{Lemma 3}}\label{Appendix C}
\noindent The divergence $D'$ shown in formula (\ref{Equation 10}) has the range \cite{Lin1991}:
\begin{equation*}
    0\leq D'(b\rightarrow b')\leq1
\end{equation*}
which means that:
\begin{equation*}
    \begin{array}{c}
    -1\leq D'(b_{H}\rightarrow\dot{\sigma}_{H})-D'(b'_{H}\rightarrow\dot{\sigma}_{H})\leq1\\
    \varrho_{\mathcal{Q}'}=[-1,1]
    \end{array}
\end{equation*}

\subsection{Proof of Theorem \ref{Theorem 1}}\label{Appendix D}
Borrowing the same two arguments we used in the proof of Lemma \ref{Lemma 1}, we obtain the \emph{dynamic} upper bound of $\mathcal{Q}''$ which corresponds to the maximum \emph{informing} flow:
\begin{equation*}
    \mathcal{Q}_{max}^{''}(\mathcal{E},b_{H}^{'})=\eta\cdot[1-\log(1+b_{H}(\sigma_{H}))]
\end{equation*}
and the \emph{dynamic} lower bound of $\mathcal{Q}''$ which corresponds to the maximum \emph{misinforming} flow:
\begin{equation*}
    \mathcal{Q}_{min}^{''}(\mathcal{E},b_{H}^{'})=-\eta\cdot\log(1+b_{H}(\sigma_{H}))
\end{equation*}
This gives us the general range of flow reported by $\mathcal{Q}$:
\begin{equation*}
    \varrho_{\mathcal{Q}''}=[-\eta\cdot\log(1+b_{H}(\sigma_{H})),\eta\cdot[1-\log(1+b_{H}(\sigma_{H}))]]
\end{equation*}

\subsection{Proof of Theorem \ref{Theorem 2}}\label{Appendix E}
\noindent Assuming a flow of $k$ bits gives us:
\begin{equation*}
    \begin{array}{l}
    \mathcal{Q}''(\mathcal{E},b_{H}^{'})=k\\
    \eta\cdot[-\log(1+b_{H}(\sigma_{H}))+\log(1+b_{H}^{'}(\sigma_{H}))]=k\\
    b_{H}^{'}(\sigma_{H})=2^{k/\eta}\cdot b_{H}(\sigma_{H})+2^{k/\eta}-1
    \end{array}
\end{equation*}

\subsection{Proof of Lemma \ref{Lemma 4}}\label{Appendix F}
\noindent \begin{equation*}
    \begin{array}{c}
    b_{1}\leq b_{2}\\
    -\log(1+b)+\log(1+b_{1})\leq-\log(1+b)+\log(1+b_{2})\\
    \mathcal{Q}''(\mathcal{E},b_{1})\leq\mathcal{Q}''(\mathcal{E},b_{2})
    \end{array}
\end{equation*}

\subsection{Proof of Theorem \ref{Theorem 3}}\label{Appendix G}
\noindent A fully uncertain attacker about the correct high state has a zero prebelief. An informing flow equal to the upper bound of $\mathcal{Q}''$:
\begin{equation*}
    \mathcal{Q}_{max}^{''}(\mathcal{E},b_{H}^{'})=\eta\cdot[1-log(1+b_{H}(\sigma_{H}))]=\eta\cdot[1-log1]=\eta
\end{equation*}
\emph{evolutes} the attacker's knowledge, and transforms her prebelief into the following postbelief:
\begin{equation*}
    b_{H}^{'}(\sigma_{H})=2^{k/\eta}\cdot b_{H}(\sigma_{H})+2^{k/\eta}-1=2^{\eta/\eta}-1=1
\end{equation*}
This postbelief captures the attacker's full certainty about the correct high state.

\subsection{Proof of Theorem \ref{Theorem 4}}\label{Appendix H}
The proof is essentially the same as the proof of Theorem \ref{Theorem 3}, although it starts by a fully certain attacker about the correct high state.

\end{document}